\documentclass[11pt]{article}

\usepackage{amsmath}
\usepackage{amssymb}
\usepackage{graphicx}
\usepackage{subfigure}

\def\middlespace {\smallskipamount=5.625pt plus1.5pt minus1.5pt
                  \medskipamount=11.25pt plus3pt minus3pt
                  \bigskipamount=22.5pt plus6pt minus6pt
                  \normalbaselineskip=22.5pt plus0pt minus0pt
                  \normallineskip=1pt
                  \normallineskiplimit=0pt
                  \jot=5.625pt
                  {\def\smallskip {\vskip\smallskipamount}}
                  {\def\medskip   {\vskip\medskipamount}}
                  {\def\bigskip   {\vskip\bigskipamount}}
                  {\setbox\strutbox=\hbox{\vrule
                    height15.75pt depth6.75pt width 0pt}}
                  \parskip 11.25pt
                  \normalbaselines}

\begin{document}

\ \vskip 1.0 in

\begin{center}
 { \Large {\bf The inevitable nonlinearity of quantum gravity falsifies}}
{\Large {\bf  the many-worlds interpretation of quantum mechanics}}

\vskip 0.2 in

\smallskip

\bigskip

\bigskip

\bigskip

{{\large
{\bf T. P. Singh\footnote{e-mail address: tpsingh@tifr.res.in} 
} \footnote{Talk given at the Meeting `Himalayan Relativity 
Dialogue', Mirik, India, 18-20 April, 2007}
}}

\medskip

{\it Tata Institute of Fundamental Research,}\\
{\it Homi Bhabha Road, Mumbai 400 005, India.}
\vskip 0.5cm
\end{center}

\vskip 1.0 in

\begin{abstract}

\noindent There are fundamental reasons as to why there should exist a
reformulation of quantum mechanics which does not refer to a classical
spacetime manifold. It follows as a consequence that quantum mechanics as we
know it is a limiting case of a more general nonlinear quantum theory, with
the nonlinearity becoming significant at the Planck mass/energy scale. This
nonlinearity is responsible for a dynamically induced collapse of the 
wave-function, during a quantum measurement, and it hence falsifies the
many-worlds interpretation of quantum mechanics. We illustrate this
conclusion using a mathematical model based on a generalized 
Doebner-Goldin equation. The non-Hermitian part of the
Hamiltonian in this norm-preserving, nonlinear, Schr\"odinger equation 
dominates during a quantum measurement, and leads to a breakdown of
linear superposition. 

\vskip 1.0 in

\centerline{\it This essay received an Honorable Mention in the}
\centerline{\it Gravity Research Foundation Essay Competition, 2007}
     
\end{abstract}

\newpage

\middlespace

\noindent 
There are two fundamental unsolved problems in our understanding of
quantum mechanics. The first is the famous problem of quantum measurement, for
which one of the possible solutions is the mechanism of decoherence, in 
conjunction with the many-worlds interpretation of quantum mechanics. An 
alternative explanation of a quantum measurement is a dynamically induced 
collapse of the wave-function, which requires modification of the 
Schr\"odinger 
equation in the measurement domain. The second unsolved fundamental problem is 
the need for a reformulation of quantum mechanics, which does not refer to a 
classical spacetime manifold \cite{singh}. In this essay we show that these 
two unsolved problems have a deep connection, and the resolution of the 
second problem implies that quantum measurement is explained by dynamically 
induced collapse of the wave-function. This, in turn, falsifies the 
many-worlds interpretation of quantum mechanics. 
 
The standard formulation of quantum theory depends on an external classical
time. The need for a reformulation of quantum mechanics which does not refer 
to a classical spacetime manifold arises because the geometry (metric and 
curvature) of the manifold is produced by {\it classical} matter fields. 
One can envisage a Universe in which there are only quantum, and no 
classical, fields. This will cause the spacetime geometry to undergo quantum 
fluctuations, which, in accordance with the Einstein hole argument, destroy 
the underlying classical spacetime manifold. However, one should still be 
able to describe quantum dynamics; hence the need for the aforementioned 
reformulation. The new formulation becomes equivalent to standard quantum 
mechanics as and when an external classical spacetime geometry becomes 
available. 

When one tries to construct such a reformulation of quantum mechanics, it
follows from very general arguments \cite{singh} that quantum gravity is 
effectively a nonlinear theory. What this means is that the 
`quantum gravitational field' acts as a source for itself. Such a nonlinearity
cannot arise in the standard canonical quantization of general
relativity, which is inherently based on linear quantum theory, and which 
leads to the Wheeler-DeWitt equation. It also follows as a consequence that 
at the Planck mass/energy scale, quantum theory itself becomes an effectively
nonlinear theory [because of self-gravity], and that the Hamiltonian 
describing a quantum system depends nonlinearly on the quantum state. The 
standard linear quantum theory is recovered as an approximation at energy 
scales much smaller than the Planck mass/energy scale. 

In \cite{singh} we have developed a model for the above-mentioned reformulation
of quantum mechanics, based on noncommutative differential 
geometry. One of the outcomes of this model is that the non-relativistic 
quantum mechanics of a particle of mass $m$ is described by a nonlinear
Schr\"odinger equation, which belongs to the Doebner-Goldin class \cite{DG} of 
nonlinear equations. The nonlinear terms depend on the mass of the
particle, and are extremely small when the
particle's mass is much smaller than Planck mass 
$m_{Pl}\sim 10^{-5}$ grams. Thus
in the microscopic domain the theory reduces to standard quantum 
mechanics. The nonlinearity becomes significant in the mesoscopic domain,
where the particle's mass is comparable to Planck mass. This is also the 
domain where the  quantum to classical transition is expected to take place; a 
nonlinearity in this domain can play a decisive role in explaining quantum
measurement. It is pertinent to mention here that current experimental tests
of quantum mechanics do not rule out such a nonlinearity, and furthermore,
because our model is based on an underlying noncommutative geometry, 
the usual objections against a nonlinear quantum mechanics do not 
apply \cite{singh}. When the particle's mass is greater than Planck mass, the
nonlinear theory reduces to standard classical mechanics. 

We now demonstrate how the Doebner-Goldin equation can explain quantum
measurement as dynamical collapse of the wave-function. The simplest
D-G equation is 
\begin{equation}
i\hbar \frac{\partial\psi}{\partial t} = -\frac{\hbar^{2}}{2m}\nabla^{2}\psi + 
V\psi +
iD(m/m_{Pl})\hbar\left(
\nabla^{2}\psi + \frac{|\nabla\psi|^{2}}{|\psi|^{2}}\psi\right).
\label{dg}
\end{equation}
The coefficient $D$ of the nonlinear, imaginary, part of the Hamiltonian
is a real constant, which depends on the ratio of the particle's mass to Planck
mass. $D$ goes to zero in the
limit $m\ll m_{Pl}$, so that then the D-G equation reduces to the linear
Schr\"odinger equation. As $m$ approaches $m_{Pl}$, $D$ becomes large enough
for the imaginary part of the Hamiltonian to dominate over the real part.
The equation is norm-preserving, although the
probability density obeys not the continuity equation, but a Fokker-Planck
equation. The equation is of interest also because it arises in the study of
unitary representations of an infinite-dimensional Lie algebra of vector 
fields $Vect(R^{3})$ and group of diffeomorphisms $Diff(R^{3})$ - these
representations provide a way to classify physically distinct quantum systems.
Further, the equation is a special case \cite{gri} of the following 
class of norm-preserving nonlinear Schr\"odinger equations
\begin{equation}
i\hbar d|\psi>/dt = H|\psi> + (1-P_\psi)U|\psi>
\label{gri}
\end{equation}
where $H$ is the Hermitian part of the Hamiltonian, $(1-P_\psi)U$ is the
non-Hermitian part, $P_\psi=|\psi><\psi|$ is the projection operator, and $U$
is an arbitrary  nonlinear operator. We will work with a  generalization
of the $U$ operator for this D-G equation, given by 
$U=iF(m/m_{Pl})\Sigma_n D_n U_n$, where
\begin{equation}
U_n =  \left[ \frac{<\psi|\nabla |\chi_n><\chi_n|\nabla|\psi>}
{<\psi|\chi_n><\chi_n|\psi>}|\chi_n><\chi_n|+\nabla^2 \right]
\label{UR}
\end{equation}
and where $D_n$ are state-dependent scalars; the real function 
$F(m/m_{Pl})$ vanishes as $m\rightarrow 0$ and monotonically increases with 
mass, and $|\chi_n>$ are a complete set of orthonormal vectors.

We will use the term `initial system' to refer to the quantum system ${\cal Q}$
on which a measurement is to be made by a classical apparatus ${\cal A}$, 
and the term `final system' to refer jointly to 
${\cal Q}$ and ${\cal A}$ after the initial system has interacted with 
${\cal A}$. A quantum measurement will be thought of as an increase in the
mass (equivalently, number of degrees of freedom) of the system, from     
the initial value $m_{\cal Q}\ll m_{Pl}$ to the final value 
$m_{\cal Q} + m_{\cal A} \gg m_{Pl}$. Clearly then, the non-Hermitian part
in (\ref{gri}), which is proportional to $U$, and hence to the scalars
$D_n$ in  (\ref{UR}), will play a critical role in the transition from the
initial system to the final system. 

We assume that ${\cal A}$ measures an observable ${\hat O}$ of ${\cal Q}$, 
having a complete set of eigenstates $|\phi_n>$. Let the quantum state of the
initial system be given as $|\psi>=\Sigma_n\ a_n|\phi_n>$. The onset of 
measurement corresponds to mapping the state $|\psi>$ to 
the state $|\psi>_F$ of the final system as
\begin{equation}
|\psi>\rightarrow |\psi>_F\ \equiv \sum_n a_n|\psi>_{Fn} = \sum_n\ a_n|\phi_n>|A_n>
\label{map}
\end{equation}
where $|A_n>$ is the state the measuring apparatus would be in, had the 
initial  system been in the state $|\phi_n>$, and the $|\chi_n>$ in
(\ref{UR}) should be understood as the direct product $|\chi_n>=|\phi_n>|A_n>$. 

During a quantum measurement the non-Hermitian part of the Hamiltonian
in (\ref{gri}) dominates over the Hermitian part, and governs the evolution
of the state $|\psi>_F$ given by (\ref{map}). Assuming that the Hermitian
operator $U_n$ maps the state $|\psi>_F$ to a state $|\xi>_{nF}$ which can be
expanded as
\begin{equation}
|\xi>_{nF}= \sum_m\ b_{nm}|\phi_m>|A_m>
\label{map2}
\end{equation}
we substitute the expansion for $|\psi>_F$ from (\ref{map}) 
in (\ref{gri}), and neglecting the Hermitian
part of the Hamiltonian we get \cite{gri}
\begin{equation}
\frac{da_n}{dt} = \frac{F(m/m_{Pl})}{\hbar}\ a_n(q_n-L)
\end{equation}
where $q_n=t_n/a_n$, $L=\Sigma_m \ a_m^{*}t_m$, $t_m=\Sigma_s b_{ms} D_s$.
If the dependence of the $D_n$'s on the state is such that the $q_n$'s are 
{\it random constants} then it follows that \cite{gri}
\begin{equation}
\frac{d}{dt}  \left(\ln \frac{a_i}{a_j}\right) = 
\frac{F(m/m_{Pl})}{\hbar}\ [q_i -q_j].
\label{evo}
\end{equation}
It follows that only the state $|\psi>_{Fi}$ having the largest real part of 
$q_i$ survives at 
the end of a measurement (since $\Sigma_n |a_n|^2=1)$, and in this manner superposition is broken. It is 
noteworthy that the time-scale for breakdown of superposition is directly proportional to Planck's constant, and it decreases with increasing mass.

The randomness of the $q_n$'s is needed to ensure that repeated measurements
of the observable ${\hat O}$ lead to different outcomes $|\psi>_{Fn}$. In
order to reproduce the observed Born probability rule, the measurement should 
cause the quantum system to collapse to the eigenstate $|\phi_n>$ with the
probability \break $p_n=|<\psi(t_0)|\phi_n>|^2$. The most plausible way to
introduce randomness in the $q_n$'s is to propose that they are related to
the random phase $\theta_0$ of the initial quantum state. As an example, if
the phase is uniformly distributed in the range $[0,2\pi]$ and the $q_n$'s are 
related to $\theta_0$ by the relations \cite{gri}
\begin{equation}
q_1=-2\pi\theta_0, \ \ 
q_n=-\frac{1}{n}\left(2\pi\theta_0 - \sum_k^{n-1}|<\psi(t_0)|\phi_k>|^2\right)
-\sum_k^{n-1} \frac{|<\psi(t_0)|\phi_k>|^2}{k}
\end{equation}
and possess the probability distribution
\begin{equation}
\omega(q_n) = |<\psi(t_0)|\phi_n>|^2\ \exp(|<\psi(t_0)|\phi_n>|^2)
\end{equation}
the Born probability rule is reproduced. 

The detailed assumptions of the above model can only be justified
after a better understanding of the relation between quantum mechanics and
noncommutative geometry has been achieved. However, it is already clear
that the natural requirement of a reformulation of quantum mechanics which
does not refer to a classical spacetime manifold compels us to consider
a nonlinear modification of the Schr\"odinger equation at the Planck 
mass/energy scale. Such a nonlinearity, which explicitly depends on
Newton's gravitational constant (via the Planck mass) is responsible for the
breakdown of superposition during a quantum measurement, and provides a 
dynamical explanation for collapse of the wave-function. 
Modifications of the Schr\"odinger equation hitherto investigated in the 
literature have been 
{\it ad} {\it hoc}, and introduced solely for the purpose of explaining 
quantum measurement. However, the nonlinear modification considered by us has 
its origin elsewhere, in quantum gravity; yet it has an impact on quantum 
measurement. 

The experimentally observed mechanism of decoherence destroys the 
{\it interference} between different
possible outcomes of measurement, but as it is based on standard linear
quantum mechanics, it preserves {\it superposition} amongst the alternatives.
In this scheme (assuming that the wave-function describes individual quantum
systems, and not merely their statistical ensemble), the only natural way to 
explain the observed lack of superposition
amongst the results of a measurement is to invoke the many-worlds 
interpretation
of quantum mechanics, wherein upon a measurement, the Universe splits into
many branches, one for every decohered state. Up until now, no theoretical
argument had been presented, to choose between a decoherence based explanation
of quantum measurement, and the  alternative explanation based on dynamically
induced collapse. Our analysis in this essay establishes that the wave-function
does collapse during a measurement, and hence the many-worlds interpretation 
stands falsified. Above all, the proposal that the initial random phase of
the quantum state is correlated with the outcome of a quantum measurement is
experimentally testable with current generation experiments, and if
confirmed, will provide the first experimental evidence for quantum gravity.

\end{document}